\documentclass[aps, pra, preprint, groupedaddress, amsfonts,
               amsmath, amssymb, showpacs, nofootinbib]{revtex4-1}
\usepackage{microtype}
\usepackage{graphicx}
\usepackage{amssymb}
\usepackage{mathtools}
\usepackage{epstopdf}
\usepackage[utf8]{inputenc}
\usepackage[T1]{fontenc}
\usepackage[usenames,dvipsnames]{xcolor}
\usepackage{hyperref}

\usepackage{color}% \textcolor{#1}{#2}

\begin{document}
\title{Independent atom model description of multiple ionization of water, methane, and ammonia molecules by proton impact}

\author{Hans J\"urgen L\"udde}
\email[]{luedde@itp.uni-frankfurt.de}
\affiliation{Center for Scientific Computing, Goethe-Universit\"at, D-60438 Frankfurt, Germany}

\author{Marko Horbatsch}
\email[]{marko@yorku.ca}
\affiliation{Department of Physics and Astronomy, York University, Toronto, Ontario M3J 1P3, Canada}

\author{Tom Kirchner}  
\email[]{tomk@yorku.ca}
\affiliation{Department of Physics and Astronomy, York University, Toronto, Ontario M3J 1P3, Canada}
\date{\today}
\begin{abstract}
We study multiple ionization in proton collisions with water, methane, and ammonia molecules using an independent-atom model. Previous work on total (net) capture and ionization cross sections is extended to treat the multiple ionization channels explicitly. We present the theoretical framework
to treat charge-state correlated processes within the independent-atom model approach which
uses the geometric screening introduced for different molecular 
geometries and orientations. Comparison of
results is made for the target molecules $\rm H_2O, CH_4, NH_3$ with an emphasis on
$q$-fold electron removal. Coincident measurements of produced molecular fragments can be used
to estimate this quantity. We find very good agreement for the model calculations for the 
water molecule, where data exist for $q=1-4$. For methane we observe reasonable agreement
with $q=1,2$, and for ammonia only for $q=1$, i.e., the experimental data show little support
for a direct multiple ionization channel in the latter case.
\end{abstract}
%
%\pacs{34.10.+x, 34.50.Gb, 34.70.+e, 36.40.-c}
%

\maketitle
\section{Introduction}
\label{intro}
The problem of electron removal from molecules impacted by charged
particles has drawn considerable interest in recent years, 
chiefly because of its relevance in diverse 
fields ranging from atmospheric chemistry to biomedical applications~\cite{Garcia12}.
In ion collisions, electron removal can be due to capture and
ionization events and both can happen at the same time, giving
rise to unstable multiply-charged cations which undergo fragmentation.
Experimental studies have looked at this problem in quite some detail, e.g.,
by measuring fragment ions in coincidence,
whereas theoretical works have usually focused on
more inclusive observables such as net
capture and ionization cross sections, which are associated with
the average number of electrons transferred from the target to
the projectile or to the continuum. 
This is so because most theoretical approaches to the problem at
hand are based on variants of effective one-electron treatments,
and net cross sections are directly accessible in such a framework.
If the projectile is singly
charged, i.e., a proton, these net cross sections have often been
equated with single-electron capture and ionization cross sections 
by virtue of interpreting
the effective one-particle calculation as a single-active electron model description.
This normally
introduces but a small error in the single- or net-electron processes studied, but
amounts to ignoring multiple ionization altogether,
and, consequently, leaves a sizable share of the accumulated
experimental data unexplained.

In earlier work concerned with molecules such as water (H$_2$O) and methane (CH$_4$), 
we have taken
a different viewpoint: We interpreted
the solutions of the single-particle Schr\"odinger equation for each initially populated molecular orbital (MO)
in an independent electron model (IEM) framework (with limited geometries) 
and combined them either multinomially
or by using a more sophisticated analysis based on determinantal
wave functions to obtain probabilities and
cross sections for multielectron processes in addition to the
net cross sections which are simply obtained by summing up the contributions
from all MOs~\cite{Murakami12a}. Similar ideas to deal with the
many-electron aspects of the problem
were used in recent classical-trajectory Monte Carlo work~\cite{Jorge2020}.

In yet another approach we applied an independent-atom-model (IAM) framework
to calculate net cross sections for ion collisions with H$_2$O, CH$_4$, and larger (bio-)molecules~\cite{hjl18, hjl19, hjl19b, hjl20, hjl20a}.
The IAM for collisions is built on the idea
that a cross section for a molecular target can be written as a
linear combination of atomic cross sections. Its simplest
incarnation is the additivity rule (labeled IAM-AR) according to which
the net cross sections for electron capture and ionization
to the continuum are simply the sums of the net cross sections
of all the atoms that make up the molecule.
In a more refined variant of the IAM weight factors are attached to each
atomic cross section in order to allow for geometric overlap in the
effective molecular cross section. We termed the latter
pixel counting method (labeled IAM-PCM), since the overlapping
cross sectional areas are calculated using pixelization.

The objective of the present work is to extend the IAM-AR and IAM-PCM
approaches in such a way that in addition to net cross sections
probabilities and cross sections
for multielectron processes can be extracted from the calculations.
This extension, which is once again based on the IEM,
is described in Sec.~\ref{sec:model}. In Sec.~\ref{sec:expt}
results of this analysis
are shown for proton collisions with H$_2$O, CH$_4$, and NH$_3$ (ammonia) which have the same number of electrons,
and are compared with experimental data and, in the case of H$_2$O,
also with our previous MO-based calculations~\cite{Murakami12a}.
The paper ends with some
concluding remarks in Sec.~\ref{sec:conclusions}.
%Atomic units, characterized by $\hbar=m_e=e=4\pi\epsilon_0=1$, are used unless otherwise stated.

\section{Theoretical considerations}
\label{sec:model}

\subsection{IEM analysis of multiple capture and ionization events}
\label{sec:iem}

We are interested in calculating probabilities and cross sections for multiple capture and ionization processes.
The starting point of the discussion is the exact expression for the probability 
of capturing $k$ and simultaneously ionizing $l$ electrons of an $N$-electron system ($k+l \le N$)~\cite{Luedde03}
\begin{equation}
P_{kl} = \binom{N}{k+l}\binom{k+l}{l}\int_{P^kI^lT^{N-k-l}}\gamma^N d^4x_1 \dots d^4x_N .
\label{eq:pkl-exact}
\end{equation}
In Eq.~(\ref{eq:pkl-exact}), $\gamma^N=\gamma^N(x_1,\ldots , x_N;t_f)$ is the $N$-particle density of the 
system taken at a final time $t_f$ long after the collision, and $\int d^4x_j$
is a short-hand for 
integration over the spatial coordinates and summation over the spin states of the $j{\rm th}$ electron, i.e., 
$x_j$ comprises space and spin coordinates.
The spatial integrals are with respect to the characteristic subspaces $P,T,I=V-P-T$, which
correspond to finding an electron bound to the projectile ($P$), the target ($T$), or
released to the ionization continuum ($I$) after the collision.

Within the framework of the IEM the
$N$-particle density is given in terms of the one-particle density matrix.
If exchange is included it reads
\begin{equation}
\gamma^N = \frac{1}{N!} 
\begin{vmatrix}
 \gamma^1(x_1,x_1) & \dots & \gamma^1(x_1,x_N)\\
 \vdots & \vdots & \vdots \\
 \gamma^1(x_N,x_1) & \dots & \gamma^1(x_N,x_N)
\end{vmatrix} ,
\end{equation}
while in Hartree approximation we have
\begin{equation}
\gamma^N = \frac{1}{N^N} \prod_{j=1}^N \gamma^1(x_j) .
\end{equation}
Using the latter, we obtain for the 
charge-state correlated probabilities 
\begin{align}
P_{kl}^{\rm IEM} &= \binom{N}{k+l}\binom{k+l}{l} \frac{1}{N^N}
 \bigg(\int_P \gamma^1(x) d^4x \bigg)^k
 \bigg(\int_I \gamma^1(x) d^4x \bigg)^l
 \bigg(\int_T \gamma^1(x) d^4x \bigg)^{N-k-l}\nonumber\\
       &= \binom{N}{k+l}\binom{k+l}{l} \frac{1}{N^N} 
	(P_C^{\rm net})^k(P_I^{\rm net})^l(P_T^{\rm net})^{N-k-l}\nonumber\\
       &= \binom{N}{k+l}\binom{k+l}{l}  
       P_C^kP_I^lP_T^{N-k-l}\nonumber\\
&= \binom{N}{k+l}\binom{k+l}{l}  
       P_C^kP_I^l(1-P_C-P_I)^{N-k-l} .
       \label{eq:Pkl}
\end{align}
In this sequence of equations, $P_x^{\rm net}$ are the (fractional) net numbers of electrons which are 
captured ($x=C$), ionized to the continuum ($x=I$), or which remain bound to
the target ($x=T$) after the collision, while $P_x=P_x^{\rm net}/N$ are 
the corresponding properly normalized probabilities, i.e., $P_x\le 1$.
Note that Eq.~(\ref{eq:Pkl}) corresponds to an IEM
analysis of capture and ionization in which shell-specific information is either
unavailable or averaged out. It has been used in a large number of ion-atom collision studies
(see, e.g., Refs.~\cite{PhysRevA.49.4556,Schultz_1990} and references therein).

\subsection{Multiple capture and ionization in ion-molecule collisions
in the IAM and IEM frameworks}

\label{sec:model1}

We are interested in
applying Eq.~(\ref{eq:Pkl}) to molecular targets: 
\begin{equation}
P_{kl}^{{\rm mol}\mid {\rm IEM}}(E,b,\phi \mid \alpha,\theta, \varphi)
      = \binom{N}{k+l}\binom{k+l}{l}  
P_{C\mid {\rm mol}}^kP_{I\mid {\rm mol}}^l(1-P_{C\mid {\rm mol}}-P_{I\mid {\rm mol}})^{N-k-l} .
\label{eq:pklmol}
\end{equation}
In Eq.~(\ref{eq:pklmol}), $E$ is the kinetic energy of the projectile (in the laboratory system), ${\bf b} = (b,\phi)$ the impact
parameter vector decomposed in polar coordinates,
and $\alpha, \theta, \varphi$ are the Euler angles for the
orientation of the target molecule in $z$-$y$-$z$ convention~\cite{Rose95}. 
Within the IAM the molecular net probabilities are given as linear combinations of their
$j=1,\ldots ,M$ atomic counterparts. In the IAM-AR one has
\begin{equation}
P_{x\mid {\rm mol}} = \frac{1}{N} \sum_{j=1}^M P_{x\mid j}^{\rm net}(E,b_j) ,
\label{eq:Pxmol-bragg}
\end{equation}
while the IAM-PCM approach yields
\begin{equation}
P_{x\mid {\rm mol}} = \frac{1}{N} \sum_{j=1}^M 
   s_{x\mid j}(E\mid \alpha, \theta, \varphi) P_{x\mid j}^{\rm net}(E,b_j) .
\label{eq:Pxmol}
\end{equation}
In Eq.~(\ref{eq:Pxmol}), 
$s_{x\mid j}$ is the PCM coefficient for capture or ionization
from the $j$th atomic constituent and
$P_{x\mid j}^{\rm net}$ [cf.~Eq.~(\ref{eq:Pxmol-bragg})]
is the corresponding net electron number for a collision with that
atom, which is located at ${\bf r}_j$ (assuming the molecule is in its ground-state configuration)
such that $b_j=\mid {\bf b} - {\bf r}_j\mid $ 
is the atomic impact parameter.

\begin{figure}
\begin{center}$
\begin{array}{cc}
\resizebox{0.425\textwidth}{!}{\includegraphics{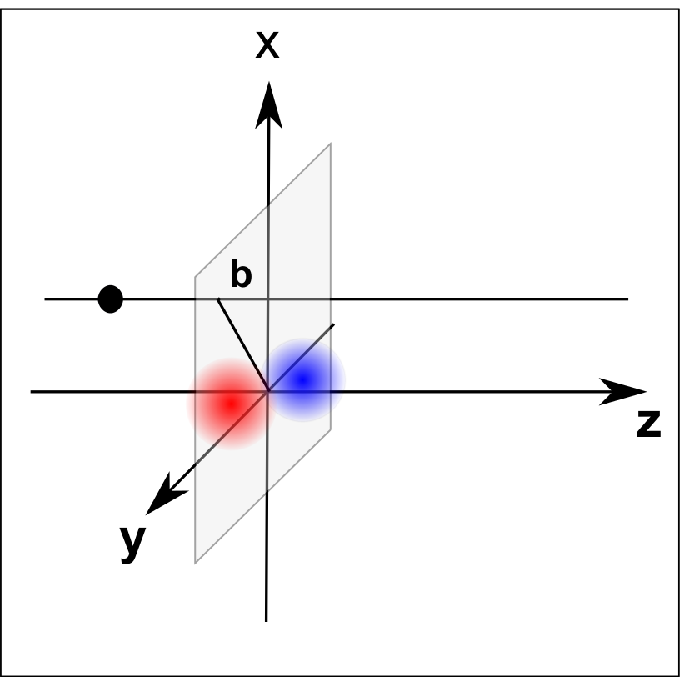}}&
\resizebox{0.55\textwidth}{!}{\includegraphics{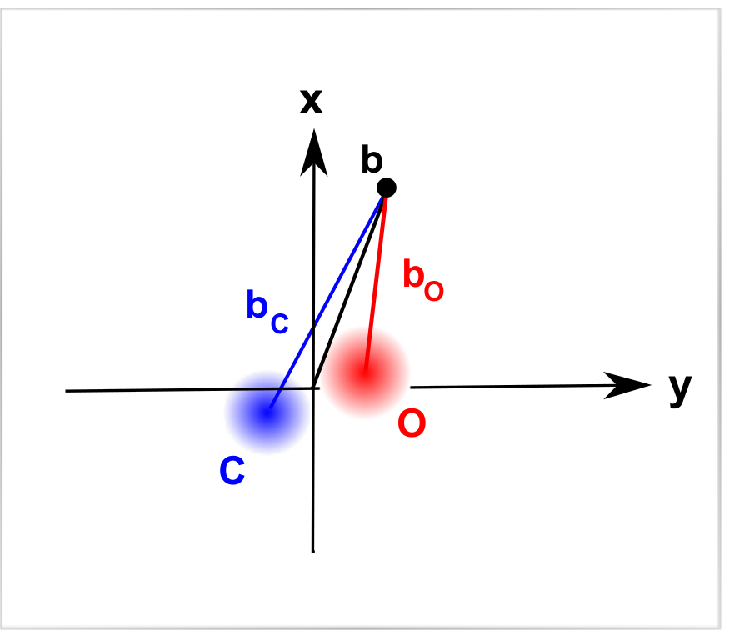}}

\end{array}$
\caption{
Scattering geometry (left panel) and projection on the impact parameter plane for a collision involving the CO molecule as an example.
}
\label{fig:fig1}
\end{center}
\end{figure}

With the Cartesian coordinates in the impact parameter plane
$x_b=b\cos \phi$ and $y_b=b\sin \phi$
one has for a given orientation of the molecule (see Fig.~\ref{fig:fig1})
$b_j = \sqrt{(x_b-x_j)^2+(y_b-y_j)^2}$. 
If the coordinates
$\bar x_j, \bar y_j, \bar z_j$ refer to a 
fixed molecular frame with respect to which the structure information
of the molecule is given,
one obtains for the coordinates in the scattering frame (i.e., the coordinate
system depicted in the left panel of Fig.~\ref{fig:fig1})
\begin{equation}
\begin{pmatrix*}[c]
 x_j \\
 y_j \\
 z_j
\end{pmatrix*} 
= 
\begin{pmatrix*}[c]
 \cos \alpha \cos \theta \cos \varphi -\sin \alpha \sin \varphi 
 & \sin \alpha \cos \theta \cos \varphi +\cos \alpha \sin \varphi
 & \sin \theta \cos \varphi\\
 -\cos \alpha \cos \theta \sin \varphi -\sin \alpha \cos \varphi 
 & \cos \alpha \cos \varphi -\sin \alpha \cos \theta \sin \varphi
 & -\sin \theta \sin \varphi\\
 -\cos \alpha \sin \theta  
 & -\sin \alpha \sin \theta  
 & \cos \theta \\  
\end{pmatrix*}
\begin{pmatrix*}[c]
 \bar x_j \\
 \bar y_j \\
 \bar z_j
\end{pmatrix*} \, .
\end{equation}

Total cross sections for a given orientation of the molecule are obtained by integration over the
impact parameter vector 
\begin{equation}
\sigma_{kl}^{{\rm mol}\mid {\rm IEM}}(E \mid \alpha,\theta, \varphi) =
\int_0^{2\pi}d \phi \int_0^\infty b db \; P_{kl}^{{\rm mol}\mid {\rm IEM}}(E,b,\phi \mid \alpha,\theta, \varphi) .
\end{equation}
Averaging over all orientations yields cross sections
\begin{equation}
\bar \sigma_{kl}^{{\rm mol}\mid {\rm IEM}}(E) = \frac{1}{8\pi^2}\int_0^{2\pi}d \varphi
   \int_0^{\pi}\sin \theta d\theta \int_0^{2\pi}d \alpha \;
\sigma_{kl}^{{\rm mol}\mid {\rm IEM}}(E \mid \alpha,\theta, \varphi) ,
\label{eq:numTCS} 
\end{equation}  
which are to be compared with experimental
data obtained for randomly oriented molecules, i.e., they are the quantities
of interest in this work.

\subsection{Inclusive probabilities}
\label{sec:model2}

The definition of the IEM probabilities ensures proper normalization
\begin{equation}
  \sum_{k,l=0}^{k+l\le N}P_{kl}^{\rm IEM} = 1.
\end{equation}
In addition, the $P_{kl}^{\rm IEM}$ 
satisfy sum rules for $q$-fold capture and ionization
\begin{align}
P_q^{C\mid {\rm IEM}} = \sum_{l=0}^{N-q} P_{ql}^{\rm IEM} , \\
P_q^{I\mid {\rm IEM}} = \sum_{k=0}^{N-q} P_{kq}^{\rm IEM} ,
\label{eq:Pq}
\end{align}
and they sum up to yield the corresponding net electron numbers
\begin{equation}
	P_x^{\rm net} = \sum_{q=1}^N qP_q^{x\mid {\rm IEM}} .
  \label{eq:sumrule2} 
\end{equation}

Following Eqs.~(31) and (32) of Ref.~\cite{Murakami12a}
we modify the probabilities and cross sections contributing to net capture according to
\begin{align}
\tilde\sigma_{1q} &= \sum_{k=1}^{N-q} \; k\;\sigma_{kq} ,
\label{eq:correct1} \\
\tilde\sigma_{k>1 q} &= 0 .
\label{eq:correct2}
\end{align}
This reinterpretation of multiple capture events as contributors
to single capture processes is a pragmatic way of dealing with the
problem that the IEM inevitably produces nonzero probabilities for
all $k+l\le N$ combinations, regardless of whether the given projectile
can accommodate $k$ electrons.
We note that double capture ($k=2$) {\it can}
occur for a proton projectile. However, it is such a rare and
highly-correlated process (associated with the formation of
a negatively-charged hydrogen ion) that in the context of the 
present IEM-based analysis it is considered unphysical and 
ruled out.

\section{Results and discussion}
\label{sec:expt}

The results reported in this work are based on the same
two-center basis generator method (TC-BGM) \cite{tcbgm} calculations
for the p+H, p+C, p+N, p+O ion-atom systems as our previously
published IAM-AR and IAM-PCM net cross sections~\cite{hjl19, hjl19b}.
In the present work we focus on IAM-PCM results, and note that they go over
into the AR limit at the highest energies, i.e., above $1 \ \rm  MeV$. 
The TC-BGM is a coupled-channel method implemented at the
IEM level of density functional theory. 
Ground-state Hartree screening and exchange effects are treated exactly, while
time-dependent
variations are known to be of minor importance
for proton collisions and are neglected. 
Further details are provided, e.g., in Ref.~\cite{hjl18}.

\subsection{p+H$_2$O collisions}
\label{sec:h2o}

We begin the discussion of results with a look at the p+H$_2$O system which we 
studied using an MO-based IEM framework in previous 
works~\cite{Murakami12a, Murakami12b, Kirchner13}.
The IAM-PCM net capture cross sections were
discussed in detail in Ref.~\cite{hjl20}, where comparison with other
calculations was also provided.

\begin{figure}
\begin{center}$
%}
%\vskip -0.5 truecm
\begin{array}{cc}
\resizebox{0.55\textwidth}{!}{\includegraphics{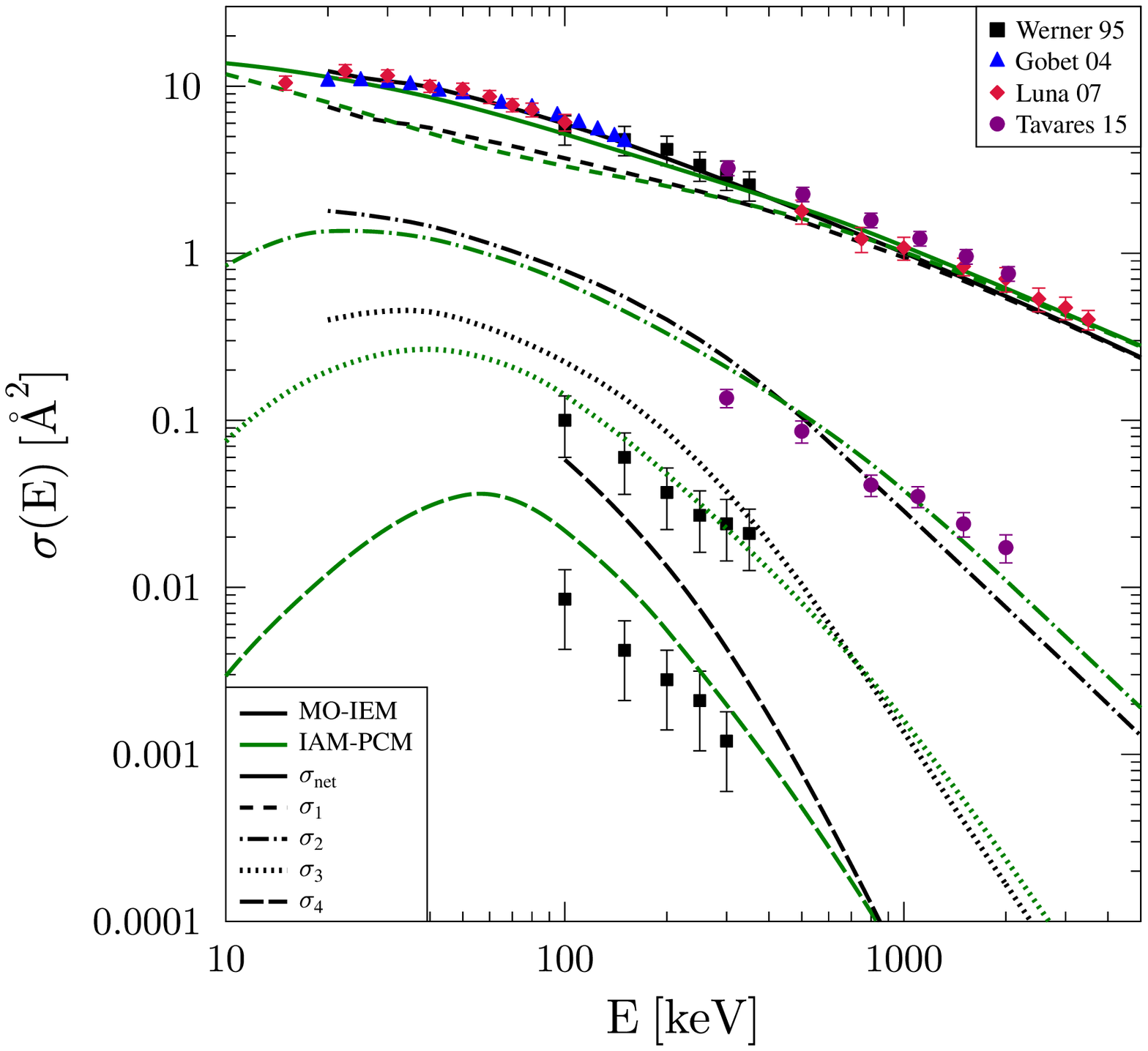}}&
\hskip -1 cm
\resizebox{0.55\textwidth}{!}{\includegraphics{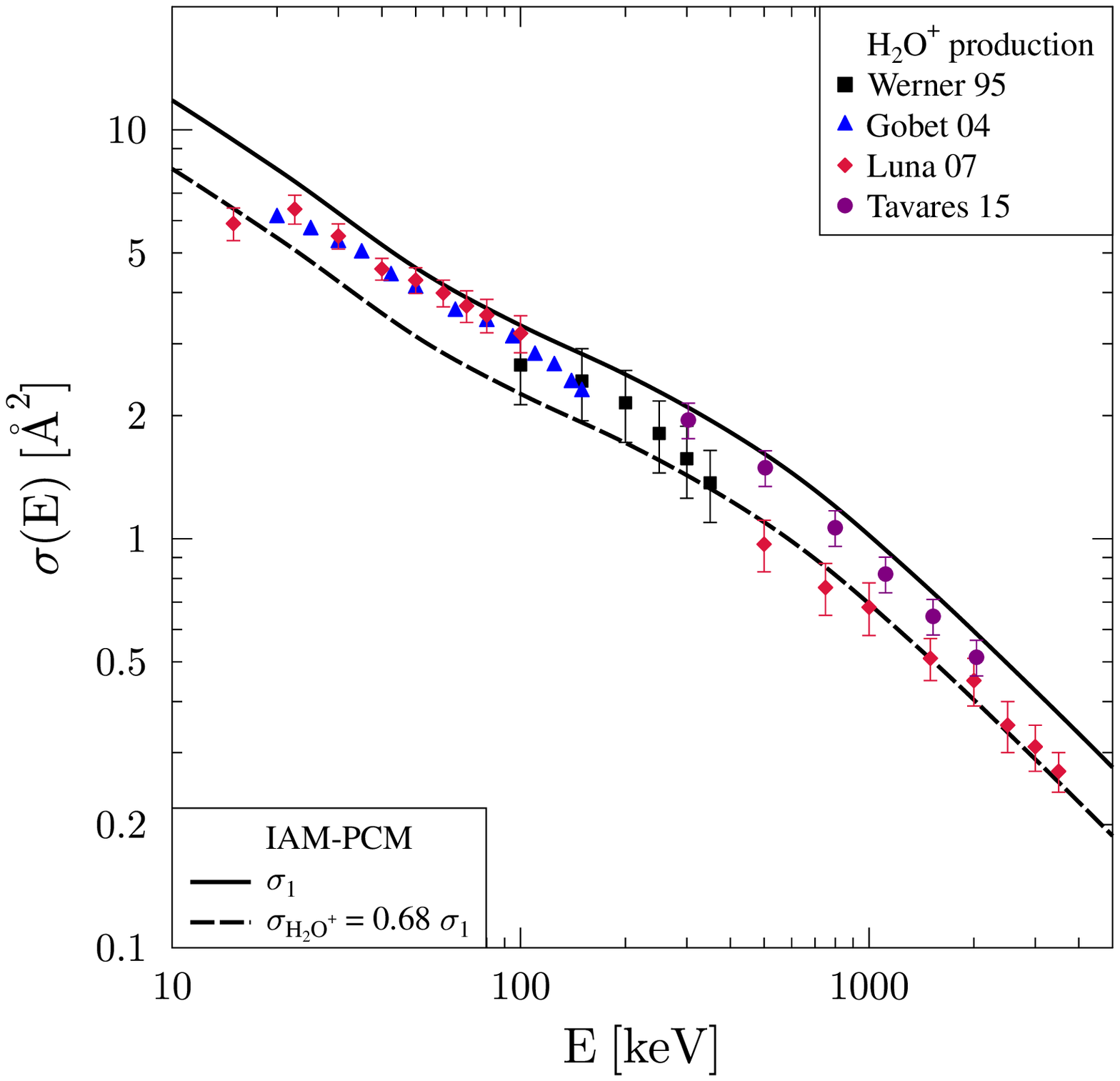}}
\end{array}$
\vskip -0.5 truecm
\caption{%
Total cross sections for proton collisions with water molecules
as functions of impact energy: in the left panel (a) for net ionization and
$q=1,2,3,4$ charge state production; in the right panel (b) the case of $q=1$ is
compared to the production of $\rm H_2O^+$.
Panel (a): 
solid curves: total (net) ionization in the MO-IEM (black, Ref.~\cite{Kirchner13}) and IAM-PCM models (green, present work) in comparison
to the fragmentation data of Ref.~\cite{Werner95} (black squares), Ref.~\cite{Gobet04} (blue triangles), Ref.~\cite{Luna07} (red diamonds), and Ref.~\cite{PhysRevA.92.032714} (magenta dots).
The broken lines are for $q=1$ (short-dashed), $q=2$ (dash-dotted), $q=3$ (dotted)
and $q=4$ (long-dashed).
The experimental data should be lower bounds for the $q$-fold electron
removal cross sections (cf. text).
Panel (b): the legend for the experimental data for the production of $\rm H_2O^+$
corresponds to those in the left panel, while the dashed line is the IAM-PCM estimate
for this channel.
}
\label{fig:p+h2o-qloss}   
\end{center}
\end{figure}

We focus on charge-state correlated cross sections in this work, but start with the
inclusive process of $q$-fold electron removal from water molecules.
Experimental data can provide mostly estimates, since the measurements
are not complete in terms of recording all coincidences. 
At intermediate energies ($100-350 \ \rm keV$), for which capture processes
play only a role at the lower end of the range multiple electron removal
cross sections for $q=1-4$ can be 
derived from ionized fragment coincidence measurements reported in 
Ref.~\cite{Werner95} (see Ref.~\cite{Murakami12b}).
Subsequent measurements at lower
energies reported singly-ionized fragment data ~\cite{Gobet04, Luna07}. 
In Ref.~\cite{Luna07} 
higher-energy data from a separate measurement were reported, 
and these were 
repeated more recently
with emphasis on
double ionization at high energies
by reporting also coincident fragment yields~\cite{PhysRevA.92.032714}.
Many of these measurements are incomplete in the sense that not all
possible reaction channels were recorded, and this is why they  provide 
typically lower bounds for $q$-fold electron removal.
One of the difficulties one is faced with concerns the absolute normalization of
the data, which in some cases is obtained by using the measured net
cross sections for total electron production 
$\sigma^- = \sigma_{\rm ion}^{\rm net}$, and
for total ion production, $\sigma^+ = \sigma_{\rm ion}^{\rm net}+\sigma_{\rm cap}^{\rm net}$~\cite{Rudd85c}.

These channels have been partially explained by the IEM calculations
of Ref.~\cite{Murakami12a, Murakami12b}, but the $q=3$ channel was found to be
overestimated already. Thus, it will be of interest to explore the success of the IAM-PCM
approach.

Other theoretical approaches for collisions with water molecules
with attempts to describe charge-state correlated cross sections include
quantum treatments using Continuum Distorted Wave (CDW) approaches~\cite{Olivera_1998, PhysRevA.93.032704, Terekhin_2018},
as well as classical-trajectory calculations with model potentials~\cite{PhysRevA.83.052704}.
These approaches have also been used to describe the ionized electron differential
cross sections, but we have not found comparisons with the $q$-fold electron removal
data of Werner {\it et al.}~\cite{Werner95} which were derived in Ref.~\cite{Murakami12b}. 
In principle, a promising approach in this respect might
be the $N$-particle classical trajectory approach~\cite{Bachi_2019}. We note, however,
that total cross sections cannot be obtained reliably from classical calculations due to their shortfall in describing ionization in distant collisions~\cite{PhysRevA.105.062822}.

For the net electron removal process experimental data are available for a wider range of energies 
in Refs~\cite{Gobet04, Luna07} and are based on fragment yields without
coincidence counting. The problem of absolute normalization does exist though,
since efficiencies for the detection of fragments are difficult to determine.
The more recent measurements of Tavares {\it et al.}~\cite{PhysRevA.92.032714}
were normalized by using a model curve that fits net recoil ion production~\cite{Rudd85c},
i.e., $\sigma^+$. In principle, such a procedure may be questioned on account of neglecting
fragments with charges $q \ge 2$, and may assign slightly too high cross sections
for the production of singly charged ions. At high energies, however, the
observed cross sections for such channels are known to be small~\cite{Werner95}.
It is worth noting that this normalization procedure affects the
$\rm H_2O^+$ channel of the data, where the results from Ref.~\cite{PhysRevA.92.032714}
are consistently higher than those of Ref.~\cite{Luna07}.

Apart from overall normalization the data of Ref.~\cite{PhysRevA.92.032714}
allow to determine the relative yields of singly ionized fragments at intermediate-to-high energies
in order to establish whether they follow the expected ratios of $68 : 16 : 13 : 3$ 
for the fragments $\rm H_2O^+$ : $\rm OH^+$ : $\rm H^+$ : $\rm O^+$  from 
photo-fragmentation~\cite{TAN1978299}. The data from Ref.~\cite{PhysRevA.92.032714} 
do support these ratios with an accuracy of a few percent. This allows one to estimate
the $q=1$ production by dividing the normalized $\rm H_2O^+$ cross section by about
$0.68$, which should at least be valid at high energies.
The data of Ref.~\cite{Luna07} were previously compared with theoretical
MO-IEM calculations in Ref.~\cite{Murakami12a, Murakami12b}.

The IAM-PCM data are presented in Fig.~\ref{fig:p+h2o-qloss}.
The curves at the top or Fig.~\ref{fig:p+h2o-qloss}(a) allow one to compare the net removal cross section $\sigma^+$ 
with the experimental fragmentation
data of Refs.~\cite{Werner95, Gobet04, Luna07, PhysRevA.92.032714}.
The most recent data at higher energies are those of Ref.~\cite{PhysRevA.92.032714}
and are on the high side. 
The data of Ref.~\cite{Werner95} were normalized by a different procedure at $E=350 \ \rm keV$,
and their net cross sections fall below these results for $\sigma^+$
for $E>200 \ \rm keV$.
Concerning the magnitude of the net ionization cross section of Ref.~\cite{Rudd85c}
we note that other quantum-mechanical theoretical works fall below it at high energies,
such as Refs.~\cite{Olivera_1998,PhysRevA.93.032704}, with the exception of 
Ref.~\cite{Terekhin_2018}.

\begin{figure}
\begin{center}
%\resizebox{0.6\textwidth}{!}{%
%}
\resizebox{0.7\textwidth}{!}{\includegraphics{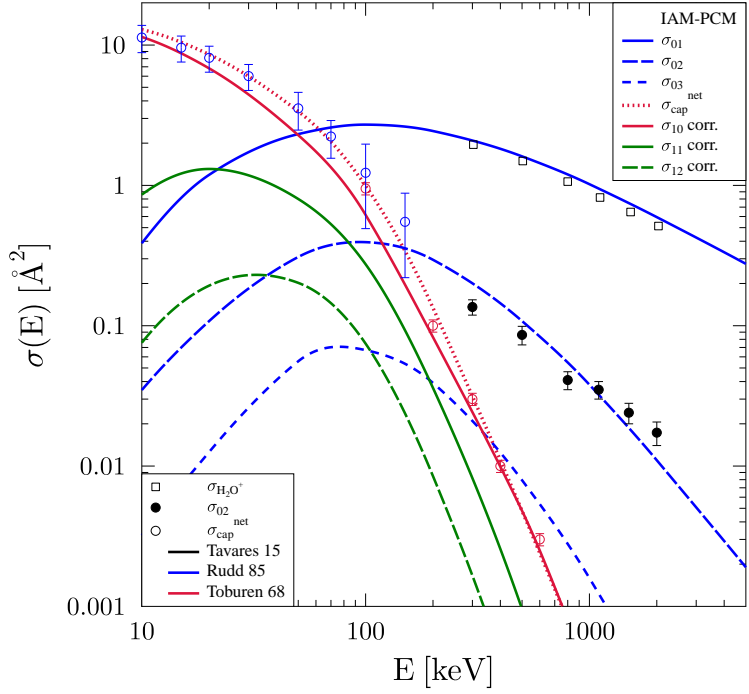}}
\vskip -0.5 truecm
\caption{%
Charge-state correlated cross sections $\sigma_{k l}$ for proton collisions with water molecules versus impact energy. IAM-PCM results are shown as curves; red solid line: pure single capture; green solid line: transfer ionization; green dashed line: transfer ionization with two electrons in the continuum;
blue solid line: pure single ionization, dashed blue line: pure double ionization, blue short-dashed line: pure triple ionization. Shown as open squares are the experimental pure single ionization data 
(exclusively) for the
${\rm H_2O^+}$ channel, and as solid dots the pure double ionization channel from Ref.~\cite{PhysRevA.92.032714}
based on coincidence measurements $\rm H^ + +OH^+$ and $\rm H^ + +O^+$,
but ignoring the  $\rm H^ + + H^+$ with neutrals channel. 
The open circles represent the derived net capture cross sections
from experiments: red open circles from Ref.~\cite{Toburen68} and the blue open circles
from Ref.~\cite{Rudd85c}. The dotted red line is the IAM-PCM net capture result~\cite{hjl20}.
}
\label{fig:p+h2o-corr}   
\end{center}
\end{figure}

It can be seen that the IAM-PCM results fall a bit short in the $\rm 30-150 \ keV$ energy
range (at about the $20 \%$ level) relative to the MO-IEM data of Ref.~\cite{Murakami12a} 
which were previously shown to
match the experimental data quite well. 
Note that the net capture cross section is described very well
within the IAM-PCM~\cite{hjl20}, as can be seen in Fig.~\ref{fig:p+h2o-corr}.

From a theoretical perspective the $q$-fold electron removal cross sections represent
an inclusive quantity, as outlined in Sect.~\ref{sec:model}. Charge-state correlated
cross sections $\sigma_{0 \, l}$ and $\tilde{\sigma}_{1 \, l}$ are calculated and then added correspondingly,
i.e.
\begin{equation}
\sigma_{q} = \tilde{\sigma}_{1 \, q-1} + \sigma_{0 \, q} \ .
\label{eq:Pqloss}
\end{equation}
The modified prescription for capture processes is used, i.e., Eq.~(\ref{eq:correct1}),
to avoid unphysical contributions from multiple electron capture which otherwise would be 
obtained in an IEM.

For the $q=1$ electron removal channel (short dashed lines in Fig.~\ref{fig:p+h2o-qloss}(a)) both the MO-IEM and IAM-PCM
methods show close agreement with each other, which is remarkable, since these
are very different calculations.
Their shape is different
from the functional shape of the net cross section.

The comparison of the $q=1$ channel with experimental data is provided separately
in Fig.\ref{fig:p+h2o-qloss}b.
The experimental data for the production of $\rm H_2O^+$ molecular ions should be a lower
bound for the $q=1$ electron removal channel. The data at low to intermediate
collision energies (\cite{Gobet04, Luna07})
agree very well with each other, and connect well to the intermediate-energy results
of Werner {\it et al.}~\cite{Werner95}. The IAM-PCM data observe this bound
very well. It is a tight bound at energies $30-100 \ \rm keV$, but we note
that the modelling of fragmentation is not straightforward in this energy regime.
The bound from the more recent higher-energy experimental data is very tight, which is
a consequence of the data being normalized to those of Ref.~\cite{Rudd85c}.
The reason for the discrepancy between the data of Ref.~\cite{Luna07}
and the data of Ref.~\cite{PhysRevA.92.032714} has not been commented upon
by the researchers, but we can assume that it is mostly related to the issue of different normalization.

At intermediate and high energies the fragmentation model which assumes
$\sigma_1 \approx 0.68 \sigma_{\rm H_2O^+}$ can be trusted, since the
fragment ratios approach constant values as a function of energy. The dashed curve
in Fig.~\ref{fig:p+h2o-qloss}b  shows that the IAM-PCM
result agrees very well with the data of Ref.~\cite{Luna07}, and is consistent
with those of Ref.~\cite{Werner95}, but less so with the more recent data of Ref.~\cite{PhysRevA.92.032714}. At energies below $100 \ \rm keV$ where the
results fall below those of Refs.~\cite{Gobet04, Luna07} the fragmentation data
are difficult to explain. We also note that a different fragmentation model has been
proposed in the literature to find better agreement with the intermediate-energy 
fragmentation data~(cf.~\cite{Olivera_1998}).

For the $q=2$ channel both theoretical models
(dash-dotted black and green lines in Fig.~\ref{fig:p+h2o-qloss}a 
agree remarkably well with each other. 
In terms of comparison with experiment for this inclusive cross section one 
has again bounds based on some coincidence data.
Werner {\it et al.}~\cite{Werner95} reported coincidence
measurements at energies of $100-350 \ \rm keV$ that involved two protons in coincidence
with neutrals or singly and doubly charged oxygen, but no coincidences between
protons and $\rm OH^+$. 
Tavares {\it et al.}~\cite{PhysRevA.92.032714}, on the other hand, for
the $300-2000 \ \rm keV$ energy range report $\rm H^+ + OH^+$ and
$\rm H^+ + O^+$ coincidences, but no $\rm H^+ + H^+$ coincidences, so their
data provide a lower bound for $q=2$ production.
Both theoretical models obey this bound for collision energies below
$1000 \ \rm keV$, and if one were to add the $\rm H^+ + H^+$ data at $300 \ \rm keV$
from Ref.~\cite{Werner95}, one would get excellent agreement, and have
an almost complete comparison (assuming $\rm O^{2+}$ production with other
neutral fragments is negligible).

At energies above $1000 \ \rm keV$ the double ionization data of 
Ref.~\cite{PhysRevA.92.032714} show a change in energy dependence
which is consistent with the appearance of an autoionization
contribution due to vacancy production in the $2a_1$ MO. We take the 
agreement of our model calculations with the experiment at lower energies
as a strong indication that such processes are not required to understand
the $q=2$ production channel unless one moves up to higher collision energies
where the $1/E^2$ fall-off of the direct double ionization process makes
these autoionization contributions visible, since they scale with the $1/E$
energy dependence of the $2a_1$ vacancy production cross section.

For the $q=3$ channel the coincidence data for the $\rm H^+ + H^+ + O^+$ 
fragment yields of Ref.~\cite{Werner95} provide a reasonable lower bound
(assuming that the unobserved  $\rm H^0 + H^+ + O^{2+}$ channel is weak).
The MO-IEM lies above the experimental data by about a factor of two, while the
IAM-PCM results represent a good match at energies below $300 \ \rm keV$. 
The last two experimental data points indicate that for $q=3$ autoionizing
processes caused by $2a_1$ vacancies may begin to contribute significantly
at this three times lower energy compared to the case of $q=2$.

For $q=4$  (solid diamonds) the $\rm H^+ + H^+ + O^{2+}$ yield
is the only relevant channel as reported in  Ref.~\cite{Werner95}.
No MO-IEM results were reported in Ref.~\cite{Murakami12a}.
The IAM-PCM result (long-dashed green line) 
overestimates the lower-energy experimental data by about a factor of two.
The energy dependence of the theoretical cross section is complicated
in this range on account of the contribution of various multi-electron processes.
It may not match the experimental fall-off very well, since autoionization contributions
may be important at even lower energies than for $q=3$
due to the smallness of the direct multiple removal contributions.
The present IAM-PCM results are closer to experiment than the
MO-IEM results (which were not shown in Ref.~\cite{Murakami12b}, but
which we include here for comparison).

Summarizing the findings reported in Fig.~\ref{fig:p+h2o-qloss} we can state
the the IAM-PCM is more successful than the MO-IEM on account of the fact
that it combines IEM calculations for proton collisions with constituent atoms 
(H and O in this case) for removal of one (or more) electrons to generate $q$-fold
electron removal from the target molecule. As we observed in our previous work
on ion collisions with multi-electron targets (such as Ne~\cite{PhysRevA.62.042704}, O~\cite{PhysRevA.61.052710} 
or Ar~\cite{PhysRevA.66.052719}) the IEM works
reasonably well for processes that involve up to one more electron than the charge
state of the projectile, i.e., two electrons in the case of proton impact. In the case
of the $\rm H_2O$ target this is verified by good agreement for the $q=2$ channel,
but overestimation for $q=3$ and failure for $q=4$ when looking at MO-IEM results.
The comparison of the present $q=1-4$ cross sections with the data of Ref.~\cite{Werner95}
shows that the energy dependence in the $100-350 \ \rm keV$ range leads to a 
different slope in the double-logarithmic presentation with an apparent steepening
of the curves as one goes through the sequence from $q=1$ to $q=4$.
It would be useful in the future to have complete coincidence measurements at
higher energies (such as those from Ref.~\cite{Werner95}) in order to verify
this behavior experimentally.

We now proceed to discuss an important objective of this work, namely the application
of Eq.~(\ref{eq:numTCS}) to calculate charge-state correlated cross sections within the
IAM-PCM. In Fig.~\ref{fig:p+h2o-corr} the data for simultaneous capture of $k=0,1$
electrons, and transfer of $l=0,1,2,3$ electrons to the continuum are presented
for proton-water collisions.
The capture channels are calculated according to ${\tilde \sigma}_{k,l}$ to avoid
the fact that the trinomial evaluation within an IEM would predict unphysical multiple
capture contributions (cf. Ref.~\cite{PhysRevA.61.052710,Murakami12a}).

The pure ionization channels $\sigma_{0l}$ can be compared to the experimental work
of Ref.~\cite{PhysRevA.92.032714}. For the process where a single electron is produced
in the continuum, while the target fragmentation products together correspond to $q=1$,
i.e., no electron capture, this cross section $\sigma_{01}$ is bounded from below
by the $\rm H_2O^+$ channel. The IAM-PCM result does obey this bound when compared
to the experiment. The earlier data from Ref.~\cite{Luna07} are lower than those
of Ref.~\cite{PhysRevA.92.032714}. at the $10-30\%$ level and would form a less tight
bound.

For the process $\sigma_{02}$, i.e., pure double ionization we can add the production
cross sections for the fragmentation channels $\rm H^+ + OH^+$ and $\rm H^+ + O^+$
shown in Fig.~4 and Table II of Ref.~\cite{PhysRevA.92.032714}. Assuming that
the production of $\rm H+H + O^{2+}$ is very small the sum should yield a tight lower
bound for this channel. Again, the IAM-PCM result agrees very closely with this bound.
We note, however, that the proton-proton coincidence data of 
Werner~{\it el al.}~\cite{Werner95} show that the measurements of Ref.~\cite{PhysRevA.92.032714}
do not provide the complete $\sigma_{02}$ cross section, since the channel
$\rm H^+ +H^+$ with neutral (or negatively charged) oxygen atoms is a significant
contributor which cannot be detected with the methodology of Ref.~\cite{PhysRevA.92.032714}.

In Fig.~4 of Ref.~\cite{PhysRevA.92.032714} comparison is also made with electron impact
data at equivalent collision velocities~\cite{PhysRevA.73.040701}. These electron data have somewhat higher cross sections. The reason for this discrepancy is not clear, i.e.,
whether it would show up as a difference in proton vs antiproton collisions with $\rm H_2O$.
In Ref.~\cite{PhysRevA.73.040701} arguments were made
that auto-dissociation of the precursor state $\rm H_2O^{2+}$ was responsible
for this strong channel. 

We note the markedly different energy dependence of the cross sections $\sigma_{01}$
and $\sigma_{02}$ at high energies. While the theoretical results for $\sigma_{01}$
and $\sigma_{02}$ energies above
$E=1 \ \rm MeV$ show $1/E$ and $1/E^2$ behavior respectively 
the experimental results indicate this for $\sigma_{01}$, but not for $\sigma_{02}$:
while the first three data points follow the theory, the three higher ones indicate
a turnover towards a weaker energy dependence. This could be caused by the
onset of a competitive double ionization process: electron removal from an inner
molecular orbital leading to further autoionization. The present theoretical model does not
include such a process. Eventually, at higher collision energies the autoionization process 
will dominate, and $\sigma_{02}$ follows the same energy dependence as $\sigma_{01}$.

Our conclusion about the production mechanism for $\sigma_{02}$ complements 
that presented in Ref.~\cite{PhysRevA.92.032714}. Based on the 
height of the cross section and its energy dependence we would argue that pure double
ionization (or $q=2$ production at these energies) can be understood as a direct
production mechanism, as discussed previously in Ref~\cite{PhysRevA.93.032704}. 
Tavares~{\it et al.}~\cite{PhysRevA.92.032714} at first
argue that the near-constancy of fragmentation ratios indicates that the channel may
be generated by single ionization followed by autoionization at all energies, but in 
the end they also conclude from the energy dependence of the cross section that 
autoionization begins to play an important role only at energies above $1 \ \rm MeV$.

For the correlated capture processes (pure single capture, transfer ionization) no direct comparison
with experiment can be provided, since the projectile charge state needs to be detected in coincidence with the fragments. 
We can use the net capture cross sections at high energies from
Ref.~\cite{Toburen68} and at lower energies from Ref.~\cite{Rudd85c} to indicate that
the sum of single capture $\sigma_{10}$ and the transfer ionization channels $\sigma_{11}$
and $\sigma_{12}$ agrees well with this net cross section (as demonstrated in Ref.~\cite{hjl20}). 
This net cross section (which
in the literature is at times denoted as $\sigma_{10}$ based on a different notation from our $\sigma_{kl}$) 
includes transfer ionization: it is formed by
taking the difference of net recoil ion production $\sigma^+$ and net electron production $\sigma^-$.
Take $\sigma_{11}$ as an example: it enters $\sigma^+$ twice on account of $q=1$ and
$\sigma^-$ once. Thus, the difference $\sigma^+ - \sigma^-$ contains $\sigma_{11}$ once,
and similar arguments hold for the higher-order transfer ionization processes.

\subsection{p+CH$_4$ collisions}
\label{sec:ch4}

\begin{figure}
\begin{center}$
%}
%\vskip -0.5 truecm
\begin{array}{cc}
\resizebox{0.55\textwidth}{!}{\includegraphics{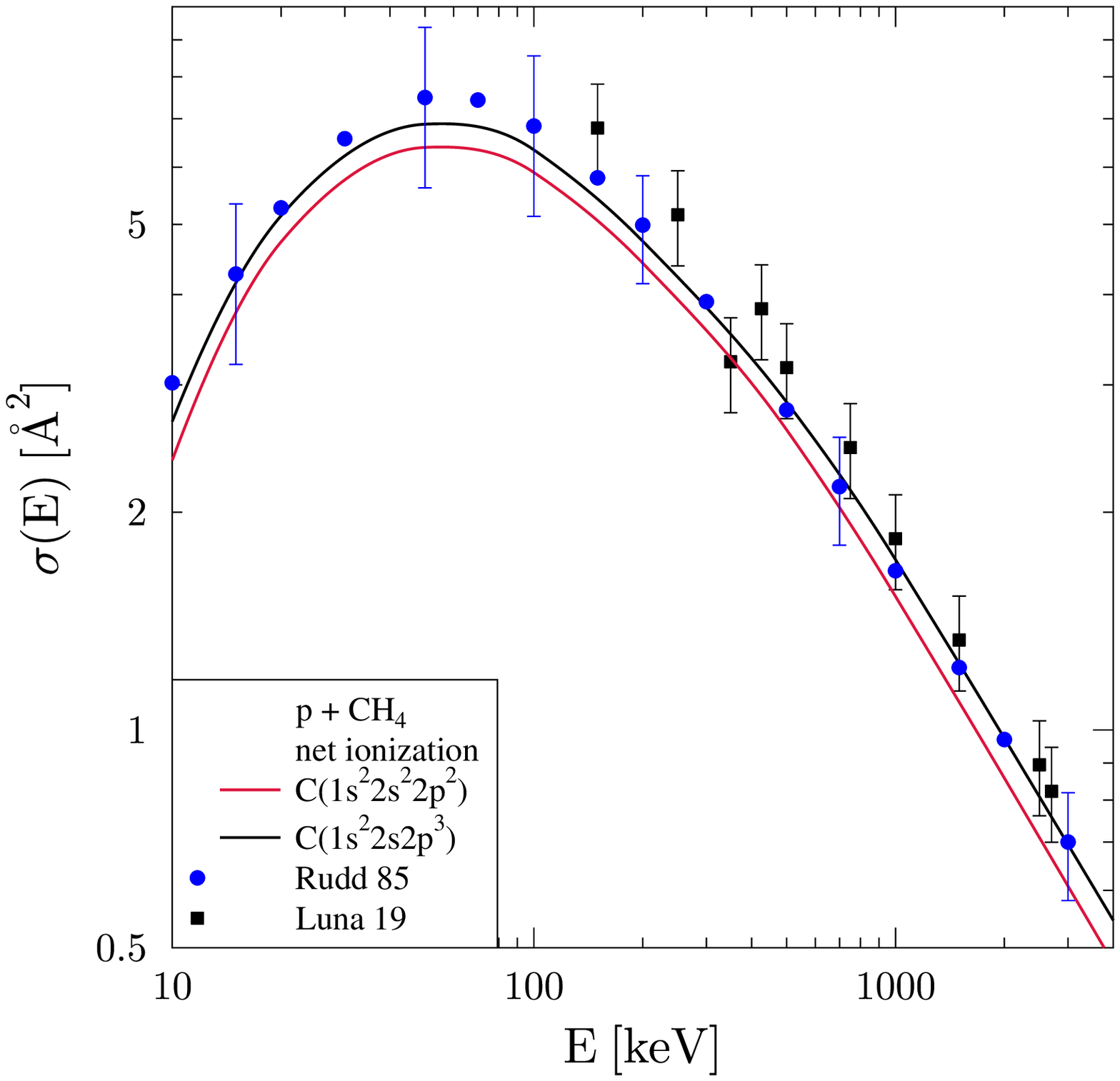}}&
\hskip -1 cm
\resizebox{0.55\textwidth}{!}{\includegraphics{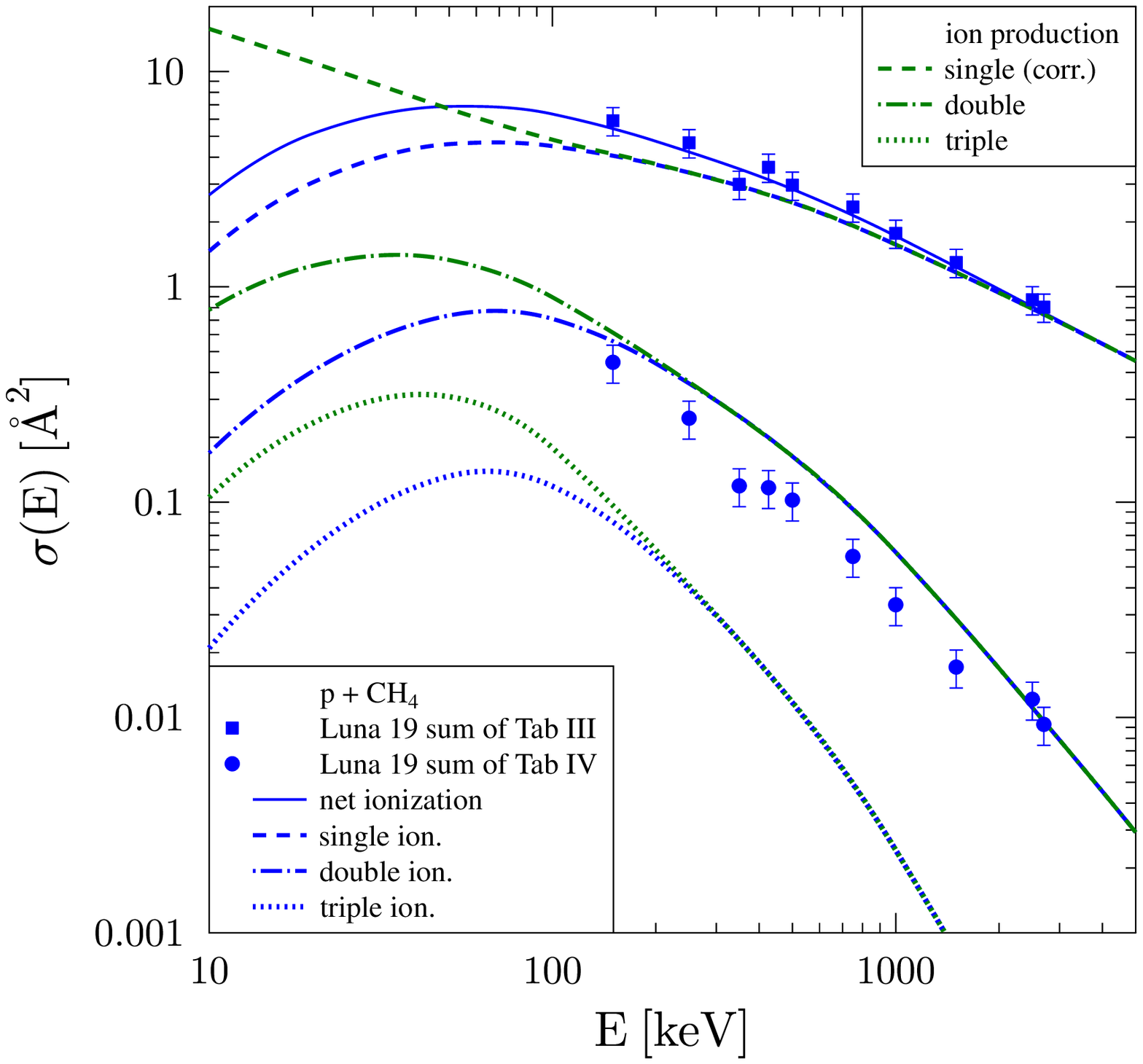}}
\end{array}$

\caption{%
(a) Net ionization cross section as a function of collision energy for proton-methane collisions.
Present IAM-PCM calculation based on ground-state carbon atom ($\rm C(1s^2 2s^2 2p^2)$) input (red line),
and based on excited carbon atom ($\rm C(1s^2 2s 2p^3)$) input (black line).
Experimental data are from Ref.~\cite{Rudd85a}, and from Ref.~\cite{Luna19} 
which were normalized to the
original measurements of Ref.~\cite{PhysRevA.28.3244}.
(b) Total cross sections for $l$-fold ionization ($\sigma_{0 l}$ shown as blue curves) 
and total ion production ($\sigma_{q}$ shown as green curves) as functions of impact energy in proton-methane collisions.
Equations~(\ref{eq:correct1})
and (\ref{eq:correct2}) are used to define the cross sections with electron
capture which enter $\sigma_{q}$ according to Eq.~(\ref{eq:Pqloss}). 
The experimental data are from Ref.~\cite{Luna19}.
The squares denote $\sigma_1 \approx \sigma_{01}$, and the solid circles 
$\sigma_2 \approx \sigma_{02}$. The IAM-PCM results
for net electron production ($\sigma^-$) are shown as a solid blue line,
$\sigma_{01}$ as a dashed blue line, $\sigma_1$ as a dashed green line, followed
by dash-dotted and dotted corresponding lines for the channels involving
$l=2,3$ and $q=2,3$ respectively.
}
\label{fig:ch4-tcskl}  
\end{center}
\end{figure}

Fig.~\ref{fig:ch4-tcskl} shows cross sections for ionization and total ion production in
proton-methane collisions.
The experimental total recoil ion production cross section for singly charged fragments
is made up of contributions from many fragments, predominantly $\rm CH_4^+$ and $\rm CH_3^+$,
 followed by smaller contributions from $\rm CH_2^+$,  $\rm CH^+$, and $\rm C^+$.
 Production of  $\rm H^+$ is comparable in size to that of $\rm CH_2^+$~\cite{Luna19}.
 Thus, the situation is more complex than for the water molecule.
 The normalization of the fragment cross sections reported by
 Luna~{\it et al.}~\cite{Luna19} is
 obtained with the help of previous coincidence measurements of 
 Ben-Itzhak~{\it et al.}~\cite{PhysRevA.49.881}, but is ultimately
 based on the original net ionization measurements of Rudd~{\it et al.}~\cite{PhysRevA.28.3244}.
 In Fig.~\ref{fig:ch4-tcskl}(a) these data (black squares) are compared to the subsequently 
 recommended values of Rudd~{\it et al.}~\cite{Rudd85a}, which are lower, but
 within error bars.

Two results are offered within the IAM-PCM approach. Previously, $\sigma^-$ and $\sigma^+$
were obtained within the method~\cite{hjl19b} based on the ground-state carbon $\rm C(1s^2 2s^2 2p^2)$
configuration.
Using an excited carbon configuration $\rm C(1s^2 2s 2p^3)$ we obtain a higher result (black vs red curve). 
This calculation reaches 
excellent agreement with the experimental data, and is used for the subsequent
analysis of multiple ionization. The justification for using an excited vs ground-state configuration
of carbon to enter the IAM-PCM analysis is based on molecular modelling:
a carbon atom cannot bind four ground-state hydrogen atoms when in the 
divalent $\rm C(1s^2 2s^2 2p^2)$ configuration. 
For a discussion of the problem we refer to Ref.~\cite{WANG2004105}.

The comparison of the solid and dashed (blue) lines in Fig.~\ref{fig:ch4-tcskl}b shows the
deviation  between $\sigma_1$ and $\sigma^-$; at energies below $E=100 \rm \ keV$
the cross sections for pure ionization fall, while the $q=1$ production cross section rises
due to capture contributions. The theoretical data are below experiment for  
low to intermediate energies, but part of the disagreement could also be a normalization
issue.

The
theoretical result for $\sigma_{02}$ (dash-dotted blue line) is bounded from below
by the experimental data of Ref.~\cite{Luna19}. These data clearly support the
notion of double ionization as a direct two-electron process, with an indication
that for $E>1000 \ \rm keV$ autoionization is becoming a competitive process.

The absence of experimental data for $q=3$ processes is a consequence of the
procedure of coincidence counting of singly charged fragments in Ref.~\cite{Luna19}.
This fact, that doubly charged fragments were not reported is also the reason why
we treat the sum of experimental channels in Table IV of  Ref.~\cite{Luna19} as
a lower bound to $\sigma_q$. The main contributions come from coincidences of
$\rm H^+$ with singly ionized hydrocarbons  and carbon atoms.
A natural channel for $q=3$ production would be $\rm H^+ + C^{2+}$,
but no such data were reported.

For $\rm CH_4$ a substantial amount of previous work, including coincidence 
measurements~\cite{PhysRevA.49.881} made it clear that a $\rm CH_4^{2+}$
channel leading to coincident charged decay products existed and had to be
modelled~\cite{Luna03}. The modelling carried out in Ref.~\cite{Luna19}
assumes that all produced $\rm CH_4^{2+}$ molecules fragment into
singly charged objects. Definitely missing in the experimental 
data for $\sigma_{02}$ are coincidences of two protons, but in contrast
to $\rm H_2O$ targets such a channel
cannot contribute a large amount since overall $\rm H^+$ production is relatively
weak contributing on the order of $5-6 \%$ only.

Capture processes (including transfer ionization) begin to play a role in ion production
($\sigma_q$) as indicated by the green curves (dashed for $q=1$, dash-dotted for
$q=2$, and dotted for $q=3$). We are not aware of experimental data that
could be used to confirm these $q=3$ results.

\subsection{p+NH$_3$ collisions}
\label{sec:nh3}

\begin{figure}
\begin{center}
%\resizebox{0.6\textwidth}{!}{%
%\begin{array}{cc}
\resizebox{0.8\textwidth}{!}{\includegraphics{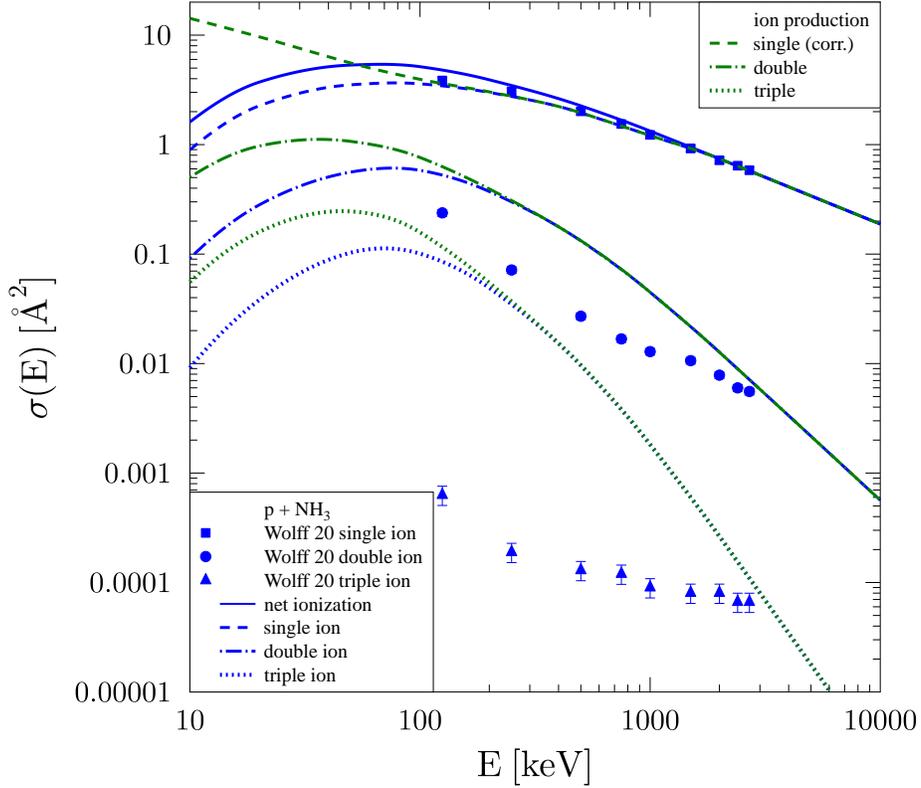}}

%\end{array}
%}
\caption{%
Total cross sections for $l$-fold ionization ($\sigma_{0 l}$) and total ion production ($\sigma_{q}$) as functions of impact energy in proton-ammonia molecule collisions.
 Equations~(\ref{eq:correct1})
and (\ref{eq:correct2}) are used to define the cross sections with electron
capture which enter $\sigma_{q}$ according to Eq.~(\ref{eq:Pqloss}). 
The experimental data are from Ref.~\cite{Wolff20}.
The squares denote $\sigma_1 \approx \sigma_{01}$, the solid circles $\sigma_2 \approx \sigma_{02}$, and triangles show $\sigma_3 \approx \sigma_{03}$ . The IAM-PCM results
for net electron production ($\sigma^-$) are shown as a solid blue line,
$\sigma_{01}$ as a dashed blue line, $\sigma_1$ as a dashed green line, followed
by dash-dotted and dotted corresponding lines for the channels involving
$l=2,3$ and $q=2,3$ respectively.
}
\label{fig:nh3-tcskl}  
\end{center}
\end{figure}

In Fig.~\ref{fig:nh3-tcskl} we compare our predictions to the data of Ref.~\cite{Wolff20}
for the ammonia target. In terms of the number of possible singly charged fragments
ammonia is much richer than water but less prolific than methane: 
the singly charged fragment yields are
strongly dominated by essentially equal amounts of $\rm NH_3^+$ and $\rm NH_2^+$.
This is followed by $\rm H^+$ production at the level of $6-7 \%$, then $\rm N^+$ 
and some $\rm H_2^+$. 
When comparing methane and ammonia one has to be aware of the fact
that the stability of decay products in these systems is different.
An interesting comparator is the observed presence of $\rm NH_3^{2+}$, for with
no doubly charged hydrocarbons being recorded experimentally
in the case of proton-methane collisions.
For the normalization of the data the recommendations 
for net cross sections from Ref.~\cite{Rudd85a} were used.

This leads to a net cross section at high energies where 
$\sigma^+ \approx \sigma^- \approx \sigma_1$, and we observe that our IAM-PCM
calculations show excellent agreement with this result over a wide range of energies. 
The situation compares well with the results for methane, for which a higher net
cross section is obtained, namely $\rm CH_4$ is ionized more effectively
than $\rm NH_3$ by about $40 \% $, even though the number of available electrons
is the same. Our model calculations show that half of this excess can be attributed
to the fact that the carbon atom enters $\rm CH_4$ in an excited configuration.

For double ionization we find a remarkable disagreement between our results
and the experimental data. The latter are obtained essentially
by summing the coincidence counts of
$\rm H^+ + NH_2^+$, $\rm H^+ + NH^+$, and $\rm H^+ + N^+$, with 
some small contributions from $\rm NH_3^{2+}$ and $\rm NH^+ + H_2^+$.
Doubly ionized nitrogen atoms were only found in coincidence with $\rm H^+$,
i.e., as part of the $q=3$ channel. It raises the question whether $\rm N^{2+}$
can be produced together with neutral hydrogen 
($\rm H_2^0 + H^0$ or $\rm H^0 + H^0+ H^0$) without being detected.

Thus, we find that compared to our prediction the experimental $q=2$ data 
show a markedly different
energy dependence: they are lower than the theoretical results by a factor of two
at $\rm 150 \ keV$ then the discrepancy increases to about a factor of four
at $\rm 400 \ keV$. At higher energies the experimental data turn around to
display an energy dependence that is comparable to the single-ionization cross
section, and they approach the theoretical values at the highest energies.

There are also experimental $q=3$ data available for the ammonia target. 
The behaviour with energy markedly disagrees with the theoretical results.
This leads to the notion that $q=2$ (and $q=3$) production does not agree
with an IEM, or the IAM, or that the measurements are incomplete and represent
lower bounds to our data in the regions where direct multiple ionization should dominate.

In order to explain their data for $\rm NH_3$ targets Wolff~{\it et al.}~\cite{Wolff20}
resort to a discussion based upon a correlated electron treatment of the molecular structure
which was used to explain electron momentum spectroscopy data
generated by electron impact~\cite{BAWAGAN1988335}. This work showed
that the details of these momentum spectra can be accounted for by going
beyond the Hartree-Fock approximation using configuration interaction methods.
On the other hand the same apparently applies to the water molecule, for which such an analysis
was reported in Ref.~\cite{BAWAGAN198719}. 
To analyze their data Wolff~{\it et al.} construct a phenomenological
fragmentation model to deal
with single ionization, while discarding multiple ionization effects.

To summarize this section we note the anomalous behavior we found for the ammonia molecule
target. For $\rm H_2O$ we find agreement with experiments that are sensitive to channels
$q=1-4$. For $\rm CH_4$ we have consistent results with experiment for $q=1,2$. Why this
should not apply to $\rm NH_3$, as well, remains a mystery at present. One way to look
into this mystery would be a study by coincidence techniques that can detect
proton-proton coincidences, such as the methodology of Ref.~\cite{Werner95}.
An additional approach would be to check the coincidence channel for electrons with
doubly charged atoms to be sensitive to processes where neutral atoms or molecules 
are produced together with doubly charged fragments.

\section{Conclusions}
\label{sec:conclusions}
We have presented a formalism for IAM calculations to evaluate cross sections for
charge-state correlated cross sections in ion-molecule collisions, 
which can then be summed to form partially inclusive cross sections, such as 
$q$-fold electron removal.

\begin{figure}
\begin{center}
\resizebox{0.8\textwidth}{!}{\includegraphics{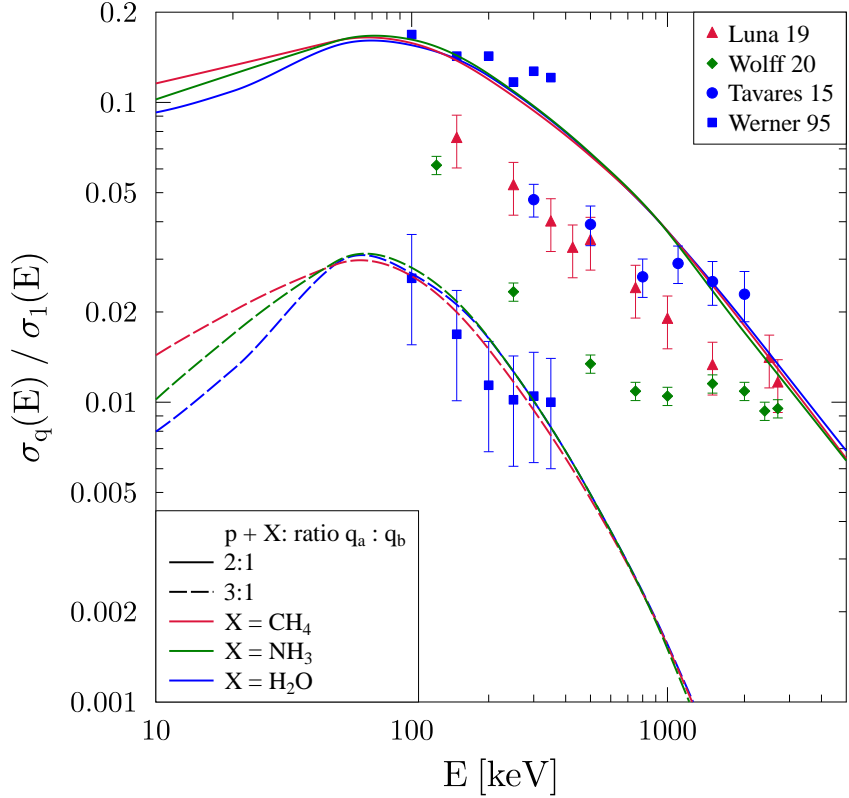}}

\caption{
Ratios of $q$-fold electron removal cross sections $\sigma_2/\sigma_1$
(solid lines)
and $\sigma_3/\sigma_1$ (dashed lines) vs energy
as calculated in the IAM-PCM approach
for proton collisions with water (blue), methane (red) and ammonia (green) collisions.
Experiments: $\rm H_2O$: Ref.~\cite{Werner95} (blue squares), Ref.~\cite{PhysRevA.92.032714} (blue dots);
$\rm CH_4$: Ref.~\cite{Luna19} (red triangles); $\rm NH_3$: Ref.~\cite{Wolff20} (green diamonds).
}
\label{fig:ratios}  
\end{center}
\end{figure}

Application of the IAM-PCM to three molecules consisting of ten electrons, where one
would expect similar results for multiple ionization led to different conclusions for the
three cases. While comparison with coincidence data for proton collisions with water
molecules agreed favorably for a multiple ionization approach, the agreement was
confirmed for methane only at the level of $q=1,2$. One might argue that the
applied coincidence technique was limited and could not detect the simultaneous
emission of two (or more) protons~\cite{Luna19}, something which was available in the work
of Werner~{\it et al.} for the water target~\cite{Werner95}.
The case of the ammonia target led, however, to a bigger puzzle, since
in this case even the direct $q=2$ multiple ionization channel turned out
to be rather weak, i.e., a shortfall by a factor of about 4-5 at intermediate energies.

From a modelling perspective this is puzzling, since the three target molecules
theoretically have similar vertical ionization energies. Using the geometry of the
neutral molecule these energies (calculated with Hartree-Fock or a correlated coupled 
cluster method~\cite{Mazziotti20}) roughly scale as $0.45 q \ \rm a.u.$ (at the $\pm 10 \%$ level
based on structure calculations for $q=0 \ldots 3$)
, and should thus be equally amenable
to an IAM approach in the high collision energy limit. 

The scaling of the IAM-PCM data is shown in Fig.~\ref{fig:ratios} in the form of ratios of
cross sections $\sigma_2/\sigma_1$ and $\sigma_3/\sigma_1$ comparing
the three molecular targets.
The theoretical results differ for the molecules at energies below $100 \ \rm keV$
only and merge into universal curves for high energies. For $\sigma_2/\sigma_1$
an energy dependence of $\ln{(E)}/E$ is observed, and as expected for 
$\sigma_3/\sigma_1 \sim (\ln{(E)}/E)^2$. Unexpected is the observation of a universal
curve for the three targets, since the constituent atoms have their own dependencies.

The experimental data for the water target (blue symbols) agree with the
finding at intermediate energies with the  $\sigma_2/\sigma_1$ 
data of Ref.~\cite{PhysRevA.92.032714} being on the low side by comparison 
at intermediate energies and then indicating a turnover to a constant at high
energies due to autoionization becoming the dominant process for $q=2$. 

The ratio $\sigma_3/\sigma_1$ agrees well with the measurements of Ref.~\cite{Werner95} .
The experimental $\sigma_2/\sigma_1$ data for methane follow the expected
$\ln{(E)}/E$ trend
over a wide range, but are lower by a factor of two, and no data are given
for $\sigma_3/\sigma_1$.

For the ammonia target the experimental results for $\sigma_2/\sigma_1$
display a very steep and counter-intuitive fall-off for the first three data points,
and then turn over to a constant. 
We do not show the experimental ratio $\sigma_3/\sigma_1$, since it would be below our
bottom scale end. The ratio is of the order of $10^{-4}$ (cf.~Fig.~\ref{fig:nh3-tcskl}) and
actually increase with $E$ for large energies.

These observations lead to the conclusion that the IAM-PCM approach is
very strong in predicting net cross sections (dominated by the $q=1$ removal
cross sections), but need to be tested on a case-by case basis for the 
higher-$q$ predictions. 
Discrete electronic excitations of the investigated molecules can also lead to 
fragmentation~\cite{Danko_2013}, and may also show different behavior among them.
Further experimental work on such molecules containing multiple hydrogen bonds, 
especially experiments capable of proton-proton coincidences are needed to 
resolve some of the remaining questions of this work.

\begin{acknowledgments}
We would like to thank the Center for Scientific Computing, University of Frankfurt for making their
High Performance Computing facilities available.
Financial support from the Natural Sciences and Engineering Research Council of Canada (NSERC) 
(RGPIN-2017-05655 and RGPIN-2019-06305) is gratefully acknowledged. \end{acknowledgments}

\bibliography{qfold}

\end{document}